\documentclass[a4paper,oneside]{article}

\usepackage{graphicx}
\usepackage{subfigure}
\usepackage{cite}
\usepackage{anysize}
\pdfoutput=1
\DeclareGraphicsExtensions{.jpg}

\begin{document}

\vskip 1cm
\marginsize{3cm}{3cm}{3cm}{1cm}

\begin{center}
{\bf{\large Numerical Studies on Time Resolution of Micro-Pattern Gaseous Detectors}}\\
~\\
Purba Bhattacharya$^a$, Nayana Majumdar$^b$, Supratik Mukhopadhyay$^b$, Sudeb Bhattacharya$^c$\\
{\em $^a$ School of Physical Sciences, National Institute of Science Education and Research, Jatni, Bhubaneswar - 752005, India}\\
{\em $^b$ Applied Nuclear Physics Division, Saha Institute of Nuclear Physics, Kolkata - 700064, India}\\
{\em $^c$ Retired Senior Professor, Applied Nuclear Physics Division, Saha Institute of Nuclear Physics, Kolkata - 700064, India}\\
~\\
~\\
~\\
~\\
~\\
{\bf{\large Abstract}}
\end{center}

The Micro-Pattern Gaseous Detectors offer
excellent spatial and temporal resolution in harsh
radiation environments of high-luminosity colliders.
In this work, an attempt has been made to establish an algorithm for
estimating the time resolution of different MPGDs.
It has been estimated numerically on the basis of
two aspects, statistics and distribution of primary electrons and their
diffusion in gas medium, while ignoring their multiplication.
The effect of detector design parameters, field configuration and the
composition of gas mixture on the resolution have also been investigated.
Finally, a modification in the numerical approach considering the threshold limit of detecting the signal has been done and tested for the RPC detector for its future implementation in case of MPGDs.

\vskip 1.5cm
\begin{flushleft}

\end{flushleft}

\vskip 1.5in
\noindent
{\bf ~$^*$Corresponding Author}: Purba Bhattacharya

E-mail: purba.bhattacharya85@gmail.com

\newpage

\section{Introduction}
\label{intro}

Owing to the use of typical manufacturing techniques for microelectronics, the
new genre of Micro-Pattern Gaseous Detectors (MPGDs) with high granularity and
very small distances between the electrodes can offer high spatial and time
 resolutions and good counting rate capability \cite{MPGD2}.
The requirement of fast collection of data in various applications of the MPGDs
  has necessitated a thorough optimization of their time resolution through the
modification of their design parameters and choice of gas mixture.
In this context, the study of the time resolution of these detectors and its
dependence on various parameters turns out to be an interesting aspect of MPGD
development for many of the current and future applications.

\begin{figure}[hbt]
\centering
\subfigure[]
{\label{Time-Micro-1}\includegraphics[scale=0.25]{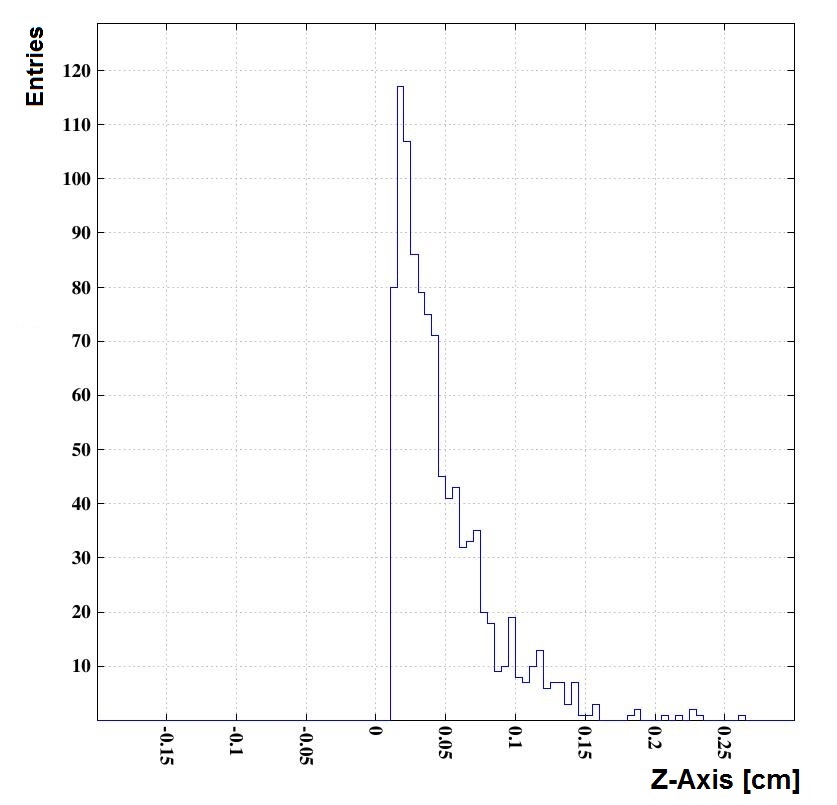}}
\subfigure[]
{\label{Time-Micro-2}\includegraphics[scale=0.25]{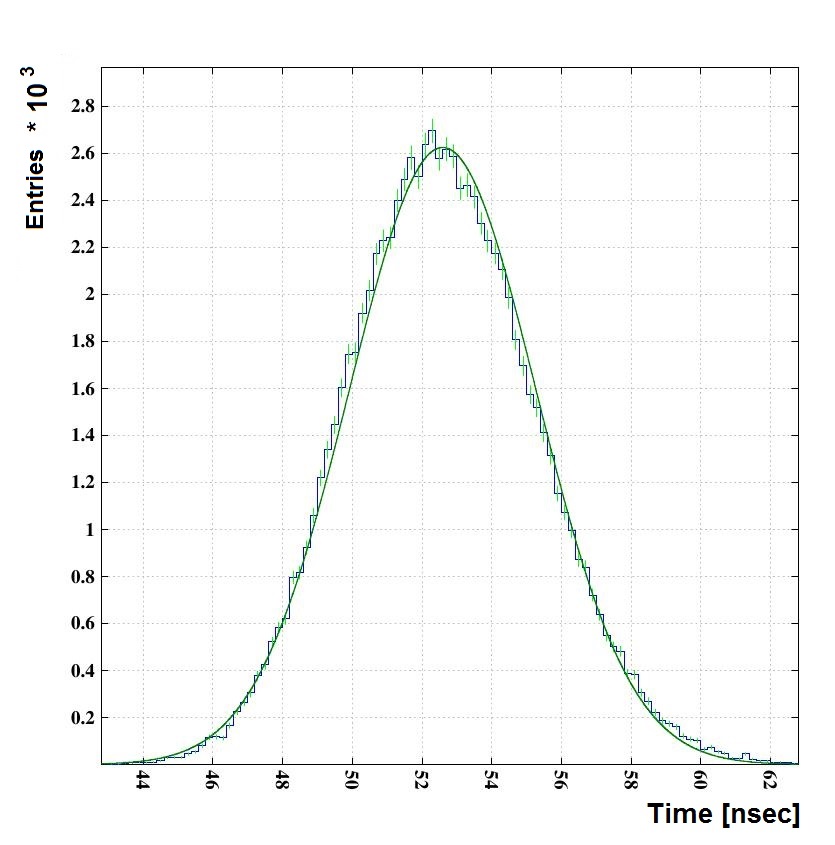}}
\subfigure[]
{\label{Final}\includegraphics[scale=0.25]{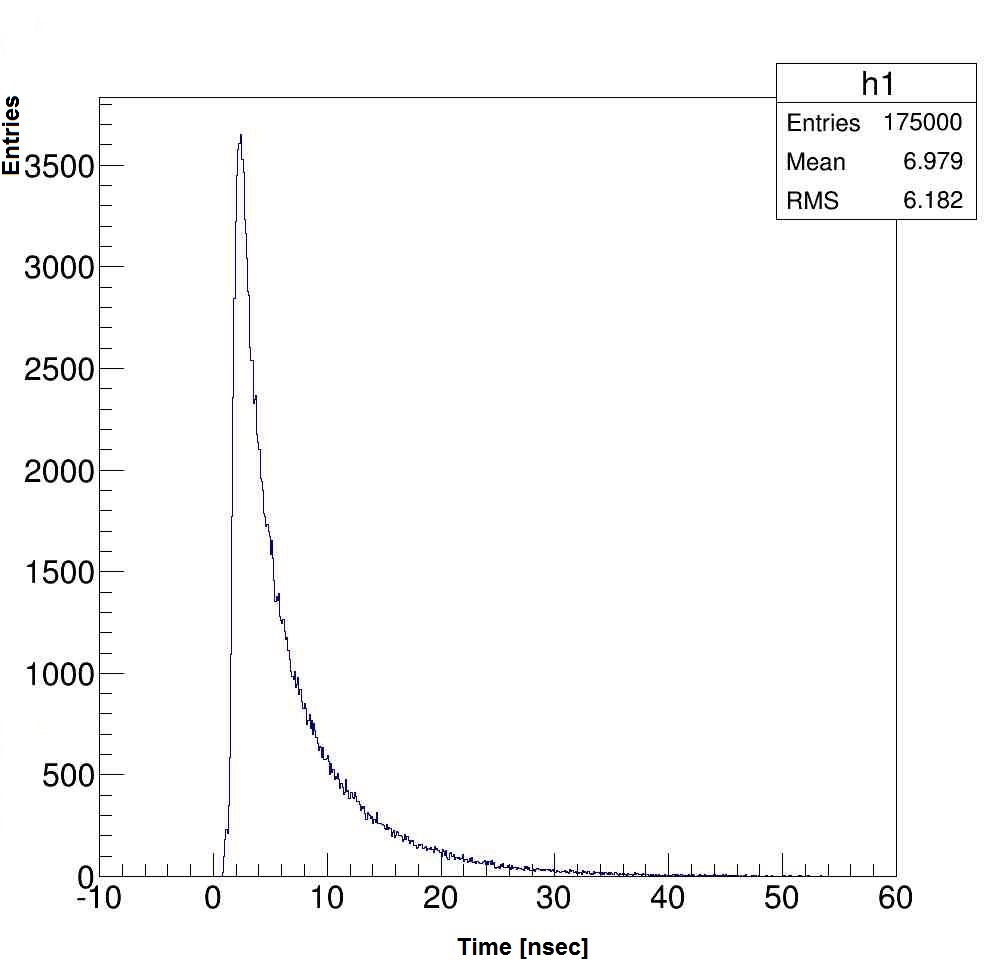}}
\caption{For a bulk Micromegas detector (a) minimum distance of an electron from the mesh plane, (b) the distribution of drift time of electron from a single point at fixed position above the mesh plane, (c) final time spectrum.}
\label{Time-Micro-Initial}
\end{figure}

The time resolution of a detector can be defined as the precision with which
the detector can distinguish between two overlapping events in terms of time.
It depends on the transit of elecrrons from their generation point to the
collecting electrode.
The spread on the duration of transit leads to a finite time resolution of the
detector \cite{Gasdetector5, Temporal1}.
The main two factors that contribute to the spread are the statistics and
distribution of the primary electrons and their diffusion in gas medium.

From event to event, the electron is not produced at the same distance from the
 read-out plane of the detector under consideration.
The general expression for the space distribution of the electron-ion pair $j$,
 closest to the read-out plane, when $\bar{N_P}$ is the average number of pairs
 produced, is obtained as follows \cite{Gasdetector5, Temporal1}:

\begin{eqnarray}
A^{\bar{N_P}}_{j}(x)~={{{~x^{j-1}}}\over{({j-1})!}}\bar{N_P}^je^{-\bar{N_P}x}
\end{eqnarray}

\noindent In particular, the distribution of the pair closer to one end of the
detection volume, is given by

\begin{eqnarray}
A^{\bar{N_P}}_{1}(x)~=~\bar{N_P}e^{-\bar{N_P}x}~=~\bar{N_P}e^{-\bar{N_P}u_et}
\end{eqnarray}

\noindent where $\mathrm{u_e}$ is the electron drift velocity.
The distribution is shown in figure~\ref{Time-Micro-1} with variance = ${{1}\over{\bar{N_P}u_e}}$.

\begin{figure}[hbt]
\centering
\subfigure[]
{\label{Time-Incl1}\includegraphics[scale=0.05]{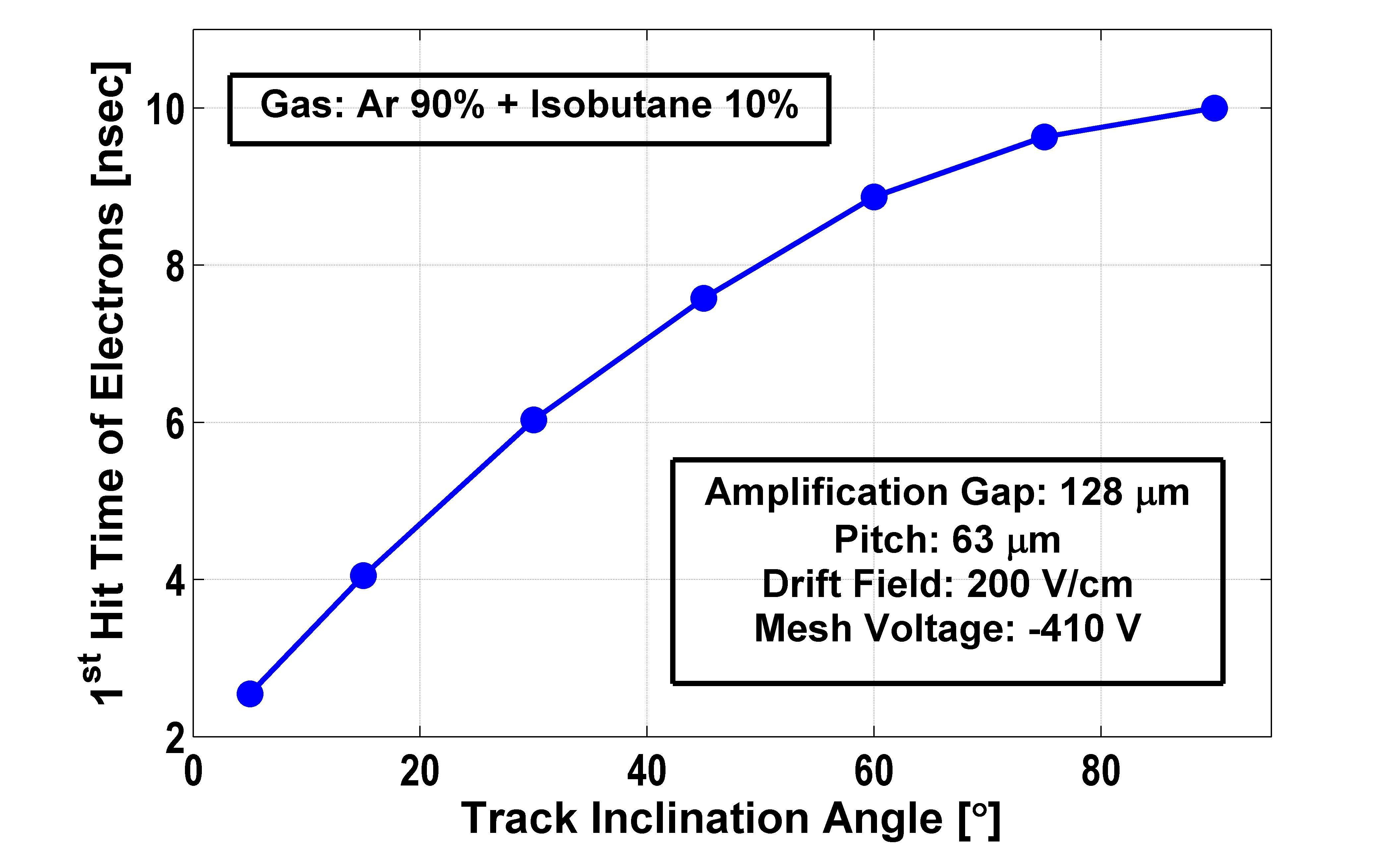}}
\subfigure[]
{\label{Time-Incl2}\includegraphics[scale=0.05]{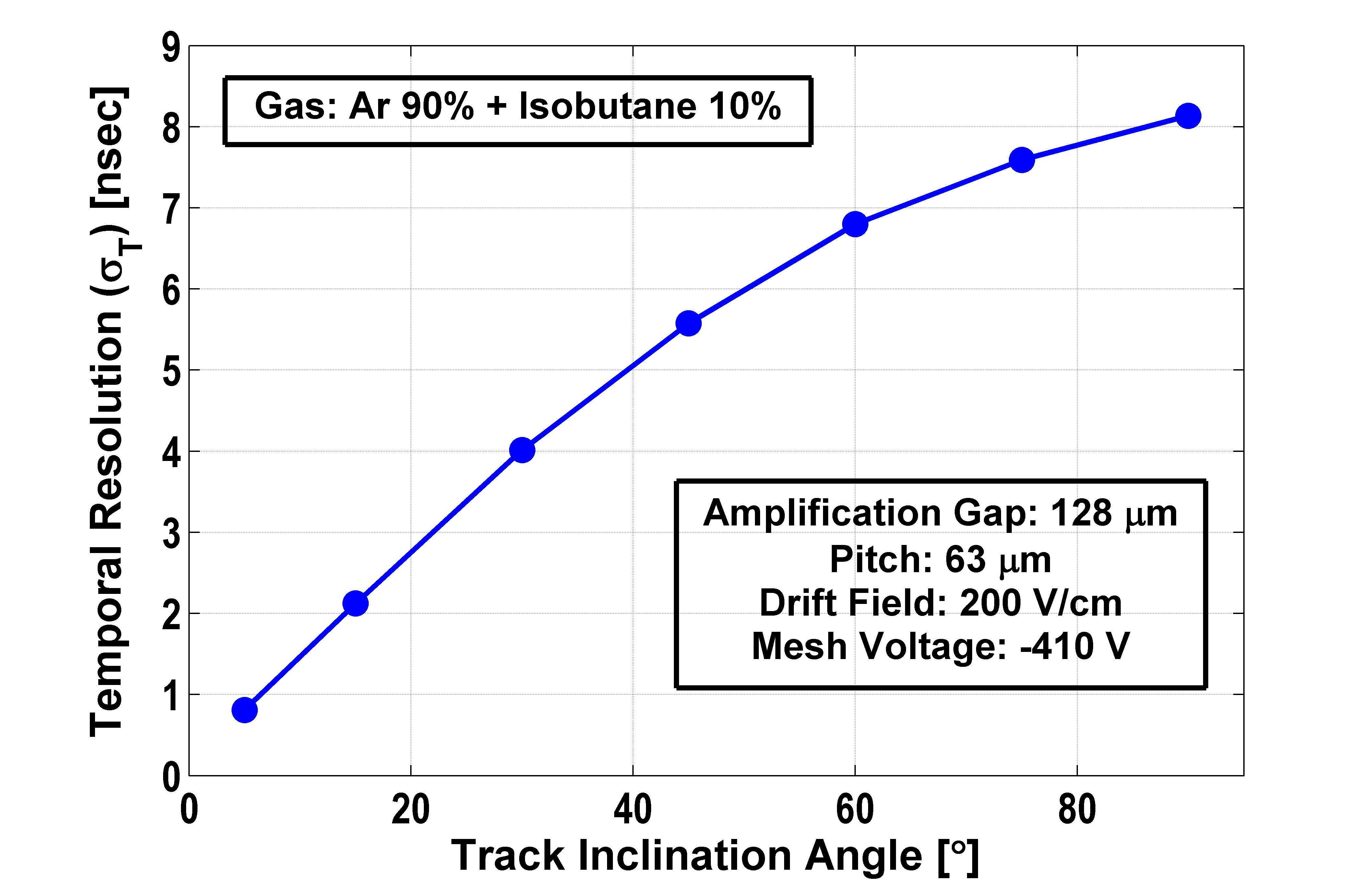}}
\caption{Variation of (a) first hit time and (b) r.m.s. with track angle in $\mathrm{Argon}$-$\mathrm{Isobutane}$ (${90:10}$) mixture for bulk Micromegas detector having amplification gap of $128~\mu\mathrm{m}$ and pitch of $63~\mu\mathrm{m}$.}
\label{Time-Incl}
\end{figure}

Again, due to the diffusion, the electrons produced at the same position in the
 gas arrive at the read-out plane at different times.
The arrival time distribution, from a particular distance is Gaussian, as shown
 in figure~\ref{Time-Micro-2}.
The mean of this distribution gives the mean arrival time and the variance is
equal to ${D}{\sqrt{z_{dist}}/u_e}$.
Here ${D}$ is the diffusion coefficient, ${z_{dist}}$ is the traveled distance.
With varying distance from the read-out plane, the mean drift time, as well as
the variance, changes accordingly.

Considering the above two factors, if the first cluster is always assumed to
generate signal that is detected by the read-out, the temporal resolution can
be defined as:

\begin{eqnarray}
\sigma_T^2~=~({{1}\over{{\bar{N_P}}{u_e}}})^2~+~({{{D}{\sqrt{z_{dist}}}}\over{{u_e}}})^2
\end{eqnarray}

In the present work, a numerical simulation of the time resolution of a few
MPGDs, is reported.
A comprehensive study on the dependence of time resolution on
detector design parameters and field configuration, has been made in addition.

\section{Simulation Tools}
\label{simtool}

The Garfield \cite{Garfield} simulation framework has been used in the following work.
The 3D electrostatic field simulation has been carried out
using neBEM \cite{neBEM} toolkit.
Besides neBEM, HEED \cite{Heed} has been used for primary
ionization calculation and Magboltz \cite{Magboltz} for computing drift, diffusion, Townsend
and attachment coefficients.

\section{Simulation Models}
\label{model}

In this work, Micromegas and GEM detectors have been opted as
two cases of the MPGD genre for studying their temporal resolution.
The design parameters of the bulk Micromegas detectors are compiled in table~\ref{Bulkdesign}, whereas that for single and triple GEM detectors are listed in table~\ref{GEMdesign}.

\begin{table}[hbt]
\caption{Design parameters of the bulk Micromegas detectors. Micromegas-wire diameter in all cases is $18~\mu\mathrm{m}$.}\label{Bulkdesign}
\begin{center}
\begin{tabular}{|c|c|}
\hline
Amplification Gap (in $\mu\mathrm{m}$) & Mesh Hole Pitch (in $\mu\mathrm{m}$)\\
\hline
$64$ & $63$ \\
\hline
$128$ & $63$ \\
\hline
$128$ & $78$ \\
\hline
$192$ & $63$ \\
\hline
\end{tabular}
\end{center}
\end{table}

\begin{table}[hbt]
\caption{Design parameters of GEM-based detectors.}\label{GEMdesign}
\begin{center}
\begin{tabular}{|c|c|}
\hline
Polymer substrate & $50~\mu\mathrm{m}$ \\
\hline
Copper coating thickness & $5~\mu\mathrm{m}$ \\
\hline
Hole diameter (copper layer) & $70~\mu\mathrm{m}$ \\
\hline
Hole diameter (Polymer substrate) & $50~\mu\mathrm{m}$ \\
\hline
Hole to hole pitch & $140~(S)$ \\
\hline
Drift Gap & $3~\mathrm{mm}$ \\
\hline
$1^{\mathrm{st}}$ Transfer gap & $1~\mathrm{mm}$ \\
\hline
$2^{\mathrm{nd}}$ Transfer gap & $2~\mathrm{mm}$  \\
\hline
Induction gap & $1~\mathrm{mm}$ \\
\hline
\end{tabular}
\end{center}
\end{table}

In our calculation, cosmic muon ($1 - 3~\mathrm{GeV}$) tracks with different
inclinations have been considered in the drift volume.
In the first approximation, the resolution has been estimated on the basis of
above two aspects while ignoring the electron multiplication.
Rather, it has been assumed that the electrons which hit the readout plane
first were multiplied adequately to produce a significant signal.
For a particular track, the drift time of those primary electrons which hit the
 readout plane first to produce a considerable signal, has been recorded.
Due the reasons mentioned above the time for the first hit varies from track to
 track and the final spectrum looks like as shown in figure~\ref{Final}, with a mean equal
to the average drift time of the first hit and r.m.s ($\sigma_{\mathrm{T}}$) equals
 to the time resolution.

\begin{table}[hbt]
\caption{Design parameters of RPC detector.}\label{RPCdesign}
\begin{center}
\begin{tabular}{|c|c|}
\hline
Bakelite thickness & $2~\mathrm{mm}$ \\
\hline
Grapite coating thickness & $20~\mu\mathrm{m}$ \\
\hline
Copper strip thickness & $200~\mu\mathrm{m}$ \\
\hline
Gas gap & $2~\mathrm{mm}$ \\
\hline
\end{tabular}
\end{center}
\end{table}

\begin{figure}[hbt]
\centering
\subfigure[]
{\label{Time-Gas1}\includegraphics[scale=0.05]{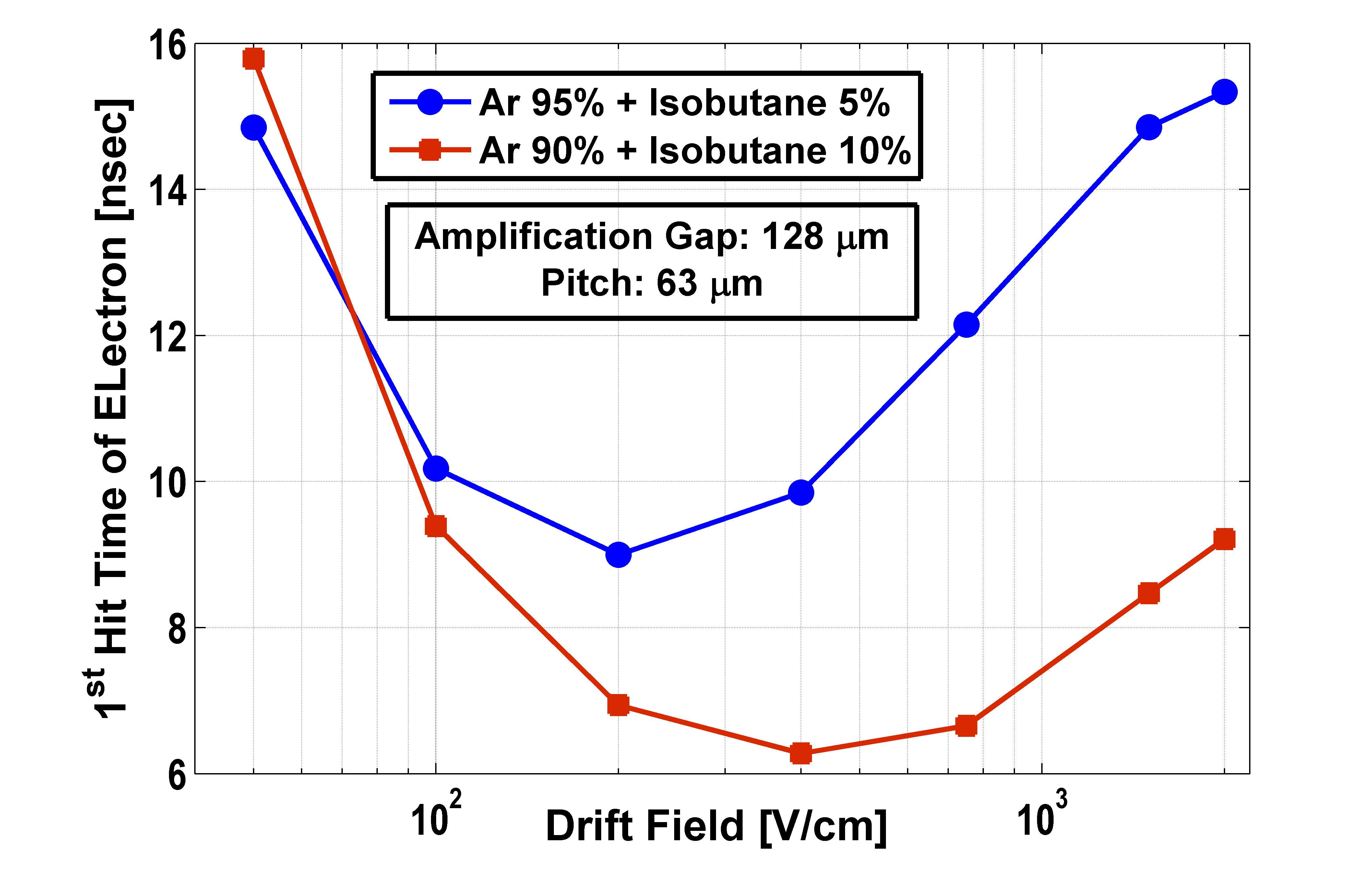}}
\subfigure[]
{\label{Time-Gas2}\includegraphics[scale=0.3]{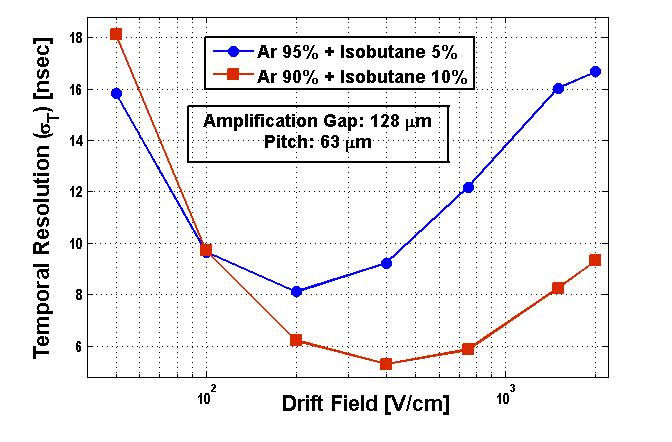}}
\subfigure[]
{\label{Time-Gas3}\includegraphics[scale=0.05]{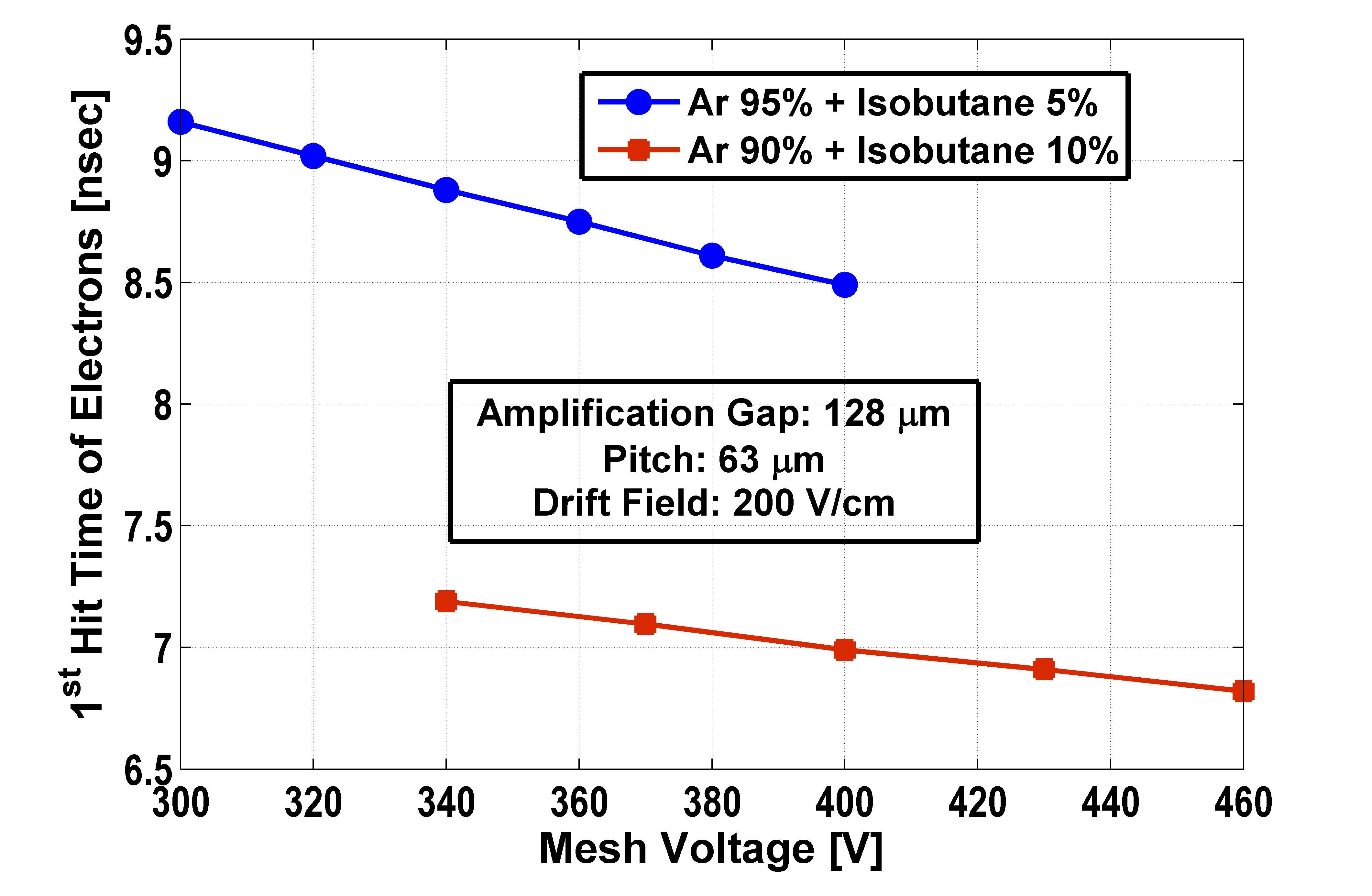}}
\subfigure[]
{\label{Time-Gas4}\includegraphics[scale=0.05]{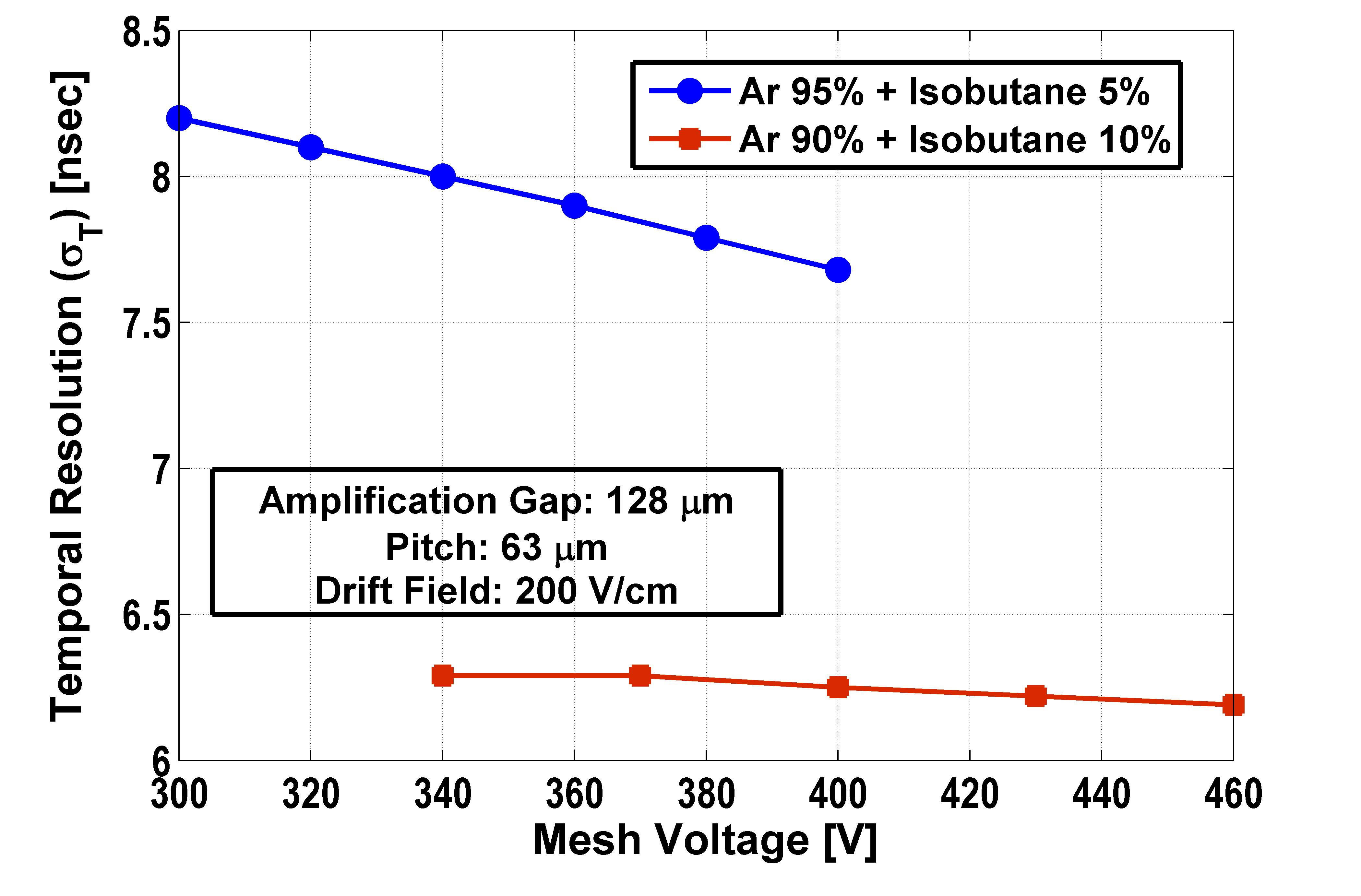}}
\caption{Variation of (a) first hit time and (b) r.m.s. with drift field in different $\mathrm{Argon}$-$\mathrm{Isobutane}$  mixtures for bulk Micromegas detector having amplification gap of $128~\mu\mathrm{m}$ and pitch of $63~\mu\mathrm{m}$. Variation of (c) first hit time and (d) r.m.s. with mesh voltage for the same detector and the same gas mixtures.}
\label{Time-Gas}
\end{figure}

\begin{figure}[hbt]
\centering
\subfigure[]
{\label{Time-Geo1}\includegraphics[scale=0.05]{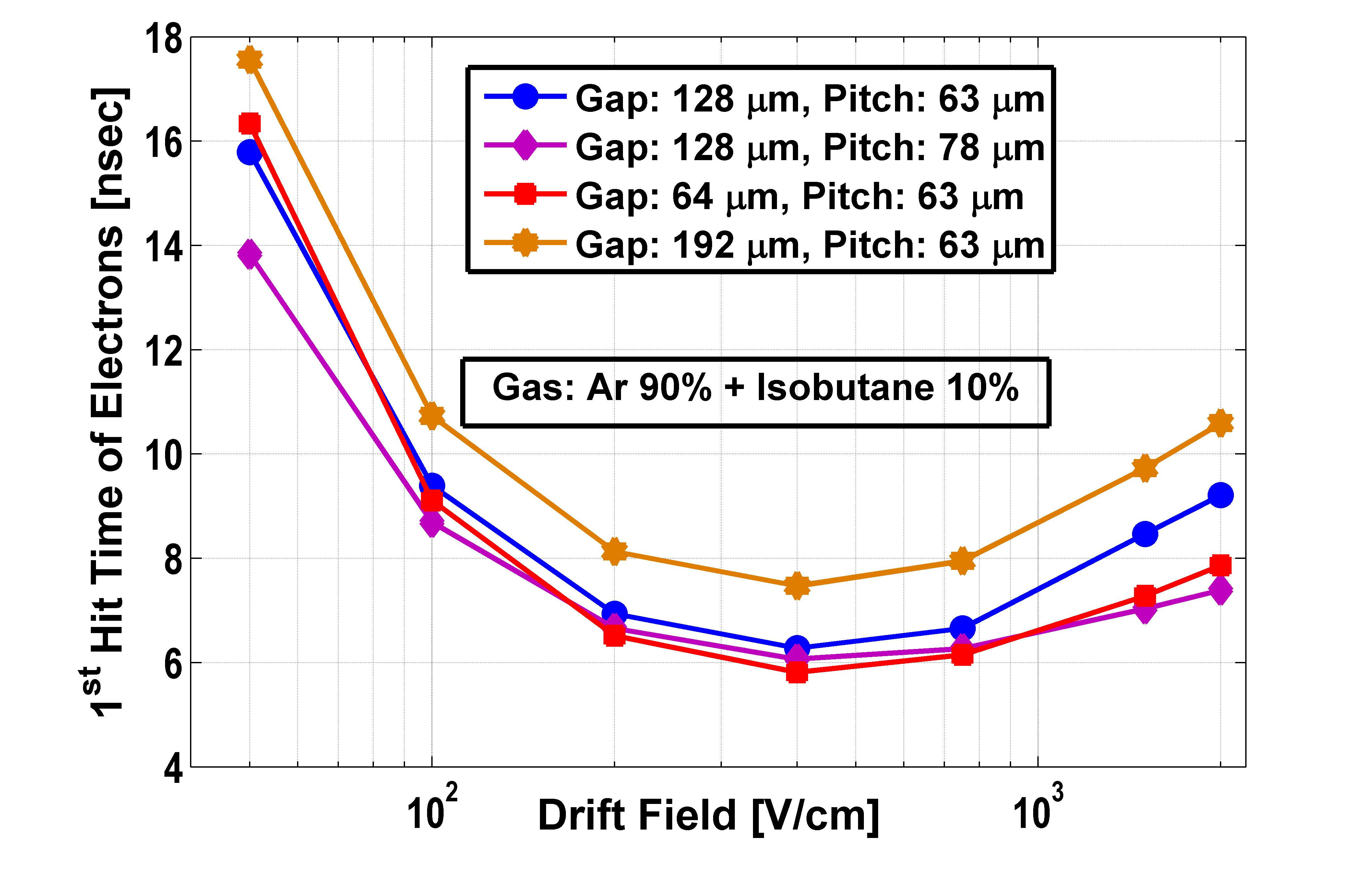}}
\subfigure[]
{\label{Time-Geo2}\includegraphics[scale=0.05]{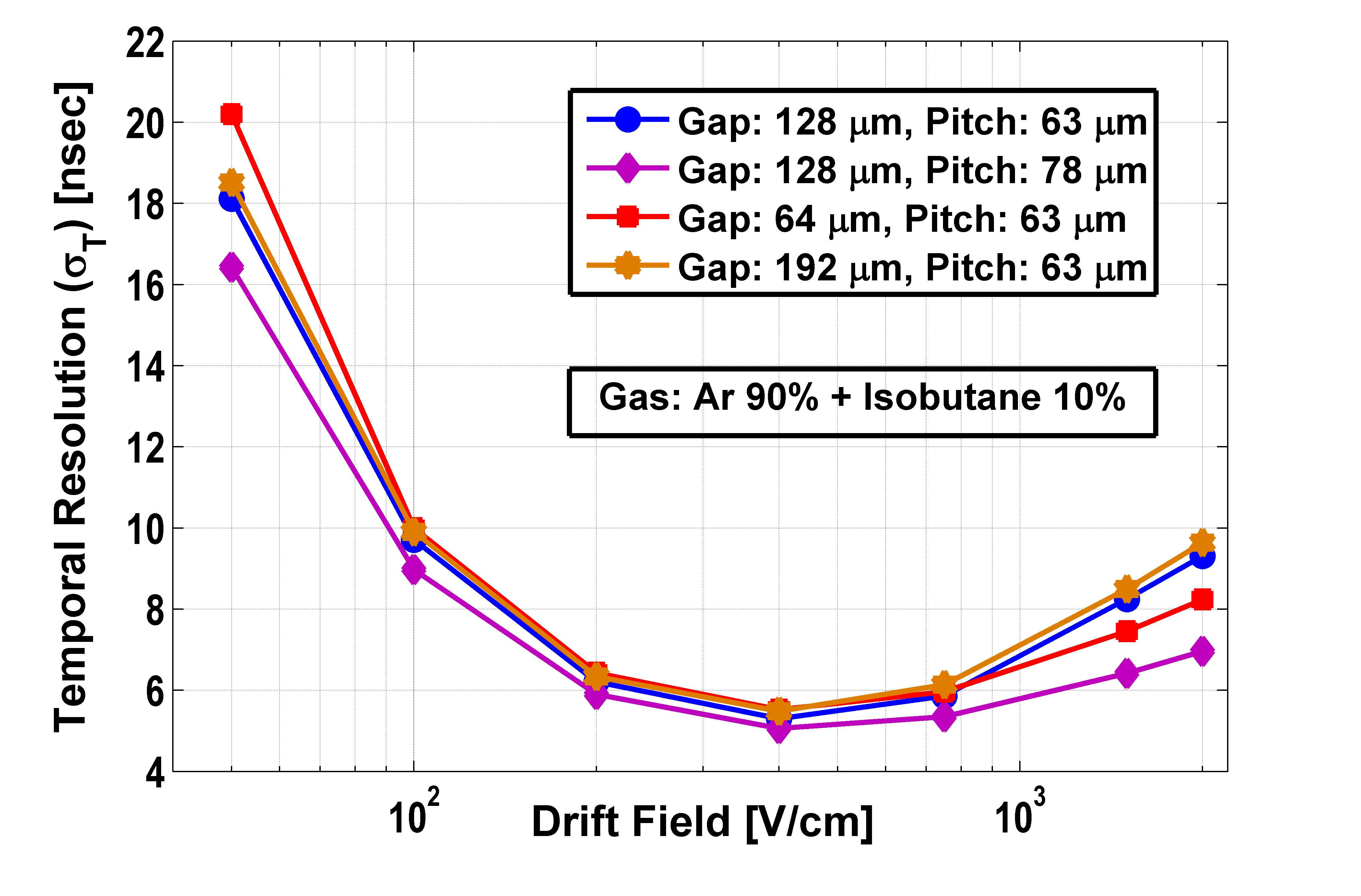}}
\caption{Variation of (a) first hit time and (b) r.m.s. for various bulk Micromegas detectors having different amplification gap and mesh hole pitch in $\mathrm{Argon}$-$\mathrm{Isobutane}$ (${90:10}$) mixture. Gap = $64~\mu\mathrm{m}$, pitch = $63~\mu\mathrm{m}$, mesh voltage = $-330~\mathrm{V}$; gap = $128~\mu\mathrm{m}$, pitch = $63~\mu\mathrm{m}$, mesh voltage = $-410~\mathrm{V}$; gap = $192~\mu\mathrm{m}$, pitch = $63~\mu\mathrm{m}$, mesh voltage = $-540~\mathrm{V}$; gap = $128~\mu\mathrm{m}$, pitch = $78~\mu\mathrm{m}$, mesh voltage = $-450~\mathrm{V}$.}
\label{Time-Geo}
\end{figure}

In our calculation, some effects of electronics, such as shaping, noise etc., have not been considered.
However, a modified simulation approach has been adopted where a threshold
limit of detecting signal has been considered which is related to the gain
variance of a detector.

\section{Results}

\subsection{Bulk Micromegas}

The temporal resolution of bulk Micromegas detector and its dependence on gas
mixture, electrostatic field configuration and geometrical parameters have been
simulated numerically \cite{Chapter6-Bulk8}.
Since the mean drift time of the first hit depends on the starting position of 
the electron, the inclination of the muon track plays an important role.
The mean drift time for different inclined tracks and the corresponding resolutions
have been plotted in figure~\ref{Time-Incl}.
The electrons from the track which makes an angle of $5^{\circ}$ with the
XY-plane, are produced close to the micromesh plane and so these electrons
traverse less path without being affected much by the diffusion.
As a result the first hit time is less and also the resolution is better.
With the increase of the inclination angle, the electrons have to travel much
longer path which causes the worsening of the resolution and larger drift time.

The variation of the first hit time and the temporal resolution with drift field have been plotted in Figure \ref{Time-Gas} for the bulk Micromegas detector having amplification gap of $128~\mu\mathrm{m}$ and pitch of $63~\mu\mathrm{m}$ in $\mathrm{Argon}$-$\mathrm{Isobutane}$ mixture with different mixing ratios.
At the lower drift field, the larger transverse diffusion is responsible for worsening of resolution and larger drift time.
At the higher drift field, due to poor funneling, the electrons traverse larger path and thus increase the drift time and temporal resolution \cite{Energy2}.
For a particular drift field, the variation of the drift time and the resolution have been plotted with the amplification field, i.e, with the mesh voltage in Figure \ref{Time-Gas3} and Figure \ref{Time-Gas4}, respectively.
The resolution is clearly improving because of better funneling and less transverse diffusion due to higher field in the amplification gap.

The effects of the variation of geometrical parameters such as variation of the amplification gap and mesh hole pitch have been also studied.
The variation of hit time and the resolution with the drift field for several bulk Micromegas detectors has been estimated (Figure \ref{Time-Geo}).
As expected, the first-hit time reduced with the reduction of amplification gap. However, no significant effect on the resolution has been observed except at higher drift fields where the detectors with larger pitch and smaller gap show comparatively better resolution.

\subsection{Single GEM}

\begin{figure}[hbt]
\centering
\subfigure[]
{\label{GEM-Hit-Drt}\includegraphics[scale=0.053]{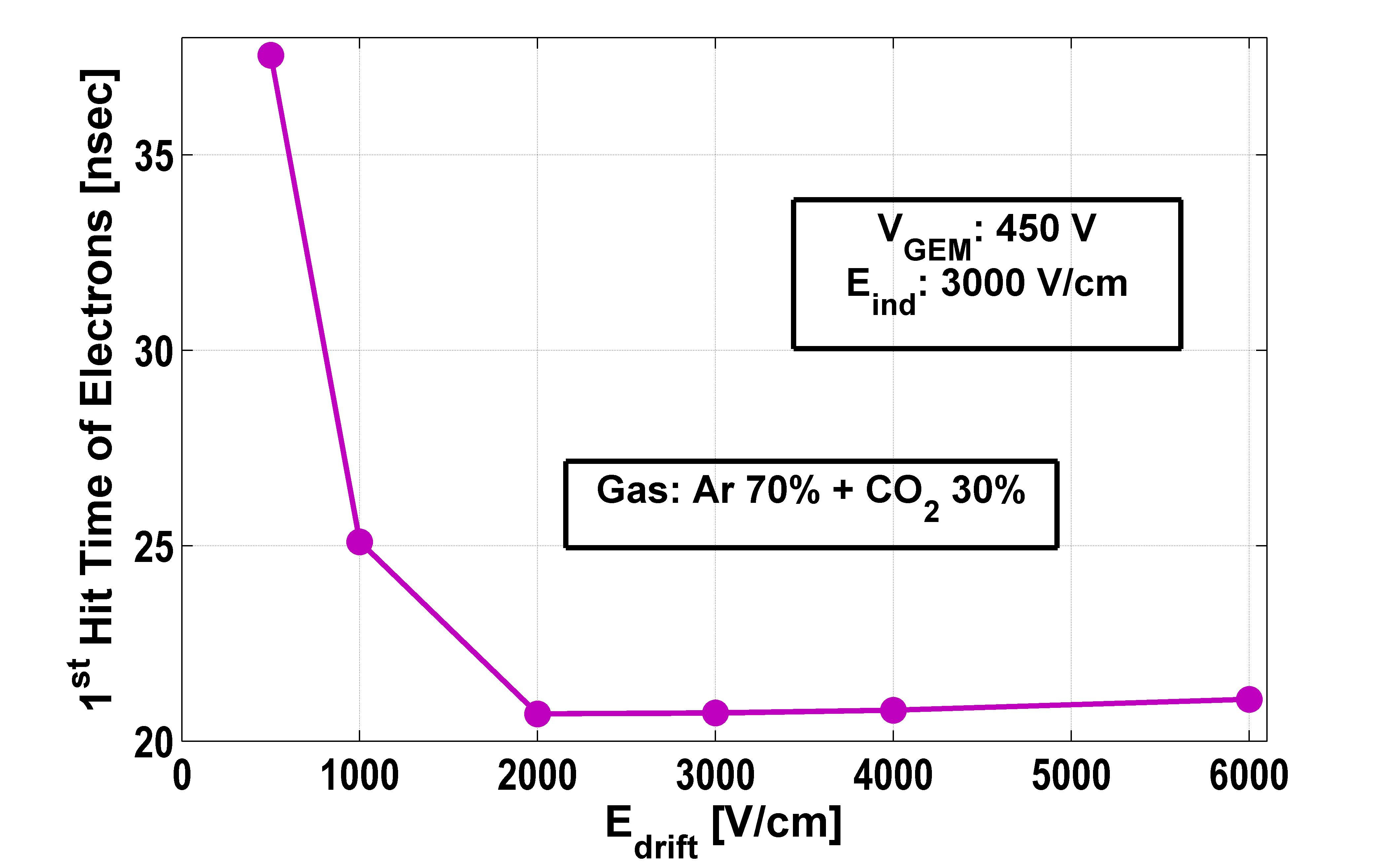}}
\subfigure[]
{\label{GEM-RMS-Drt}\includegraphics[scale=0.05]{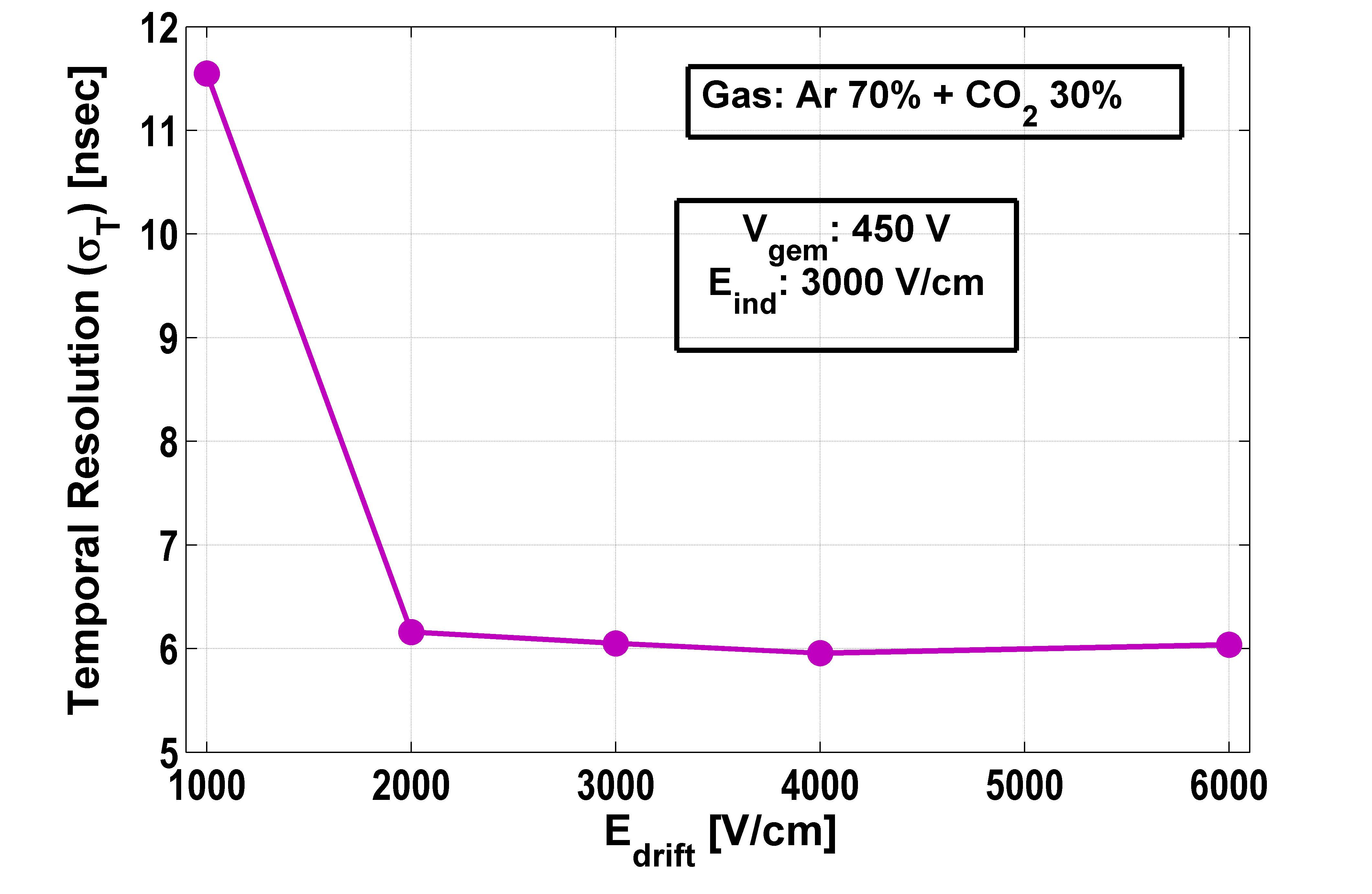}}
\subfigure[]
{\label{GEM-Hit-Vgem}\includegraphics[scale=0.05]{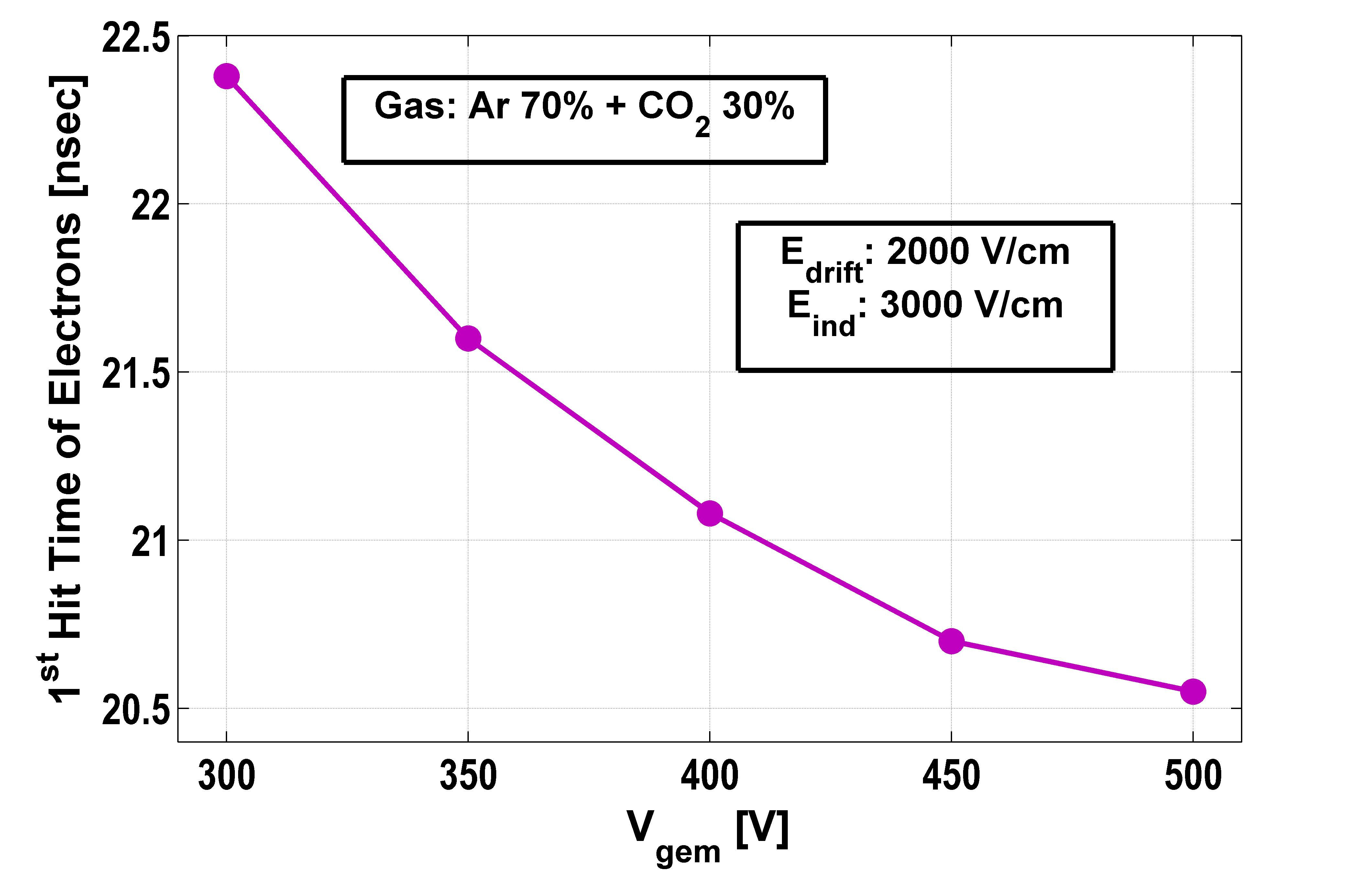}}
\subfigure[]
{\label{GEM-RMS-Vgem}\includegraphics[scale=0.05]{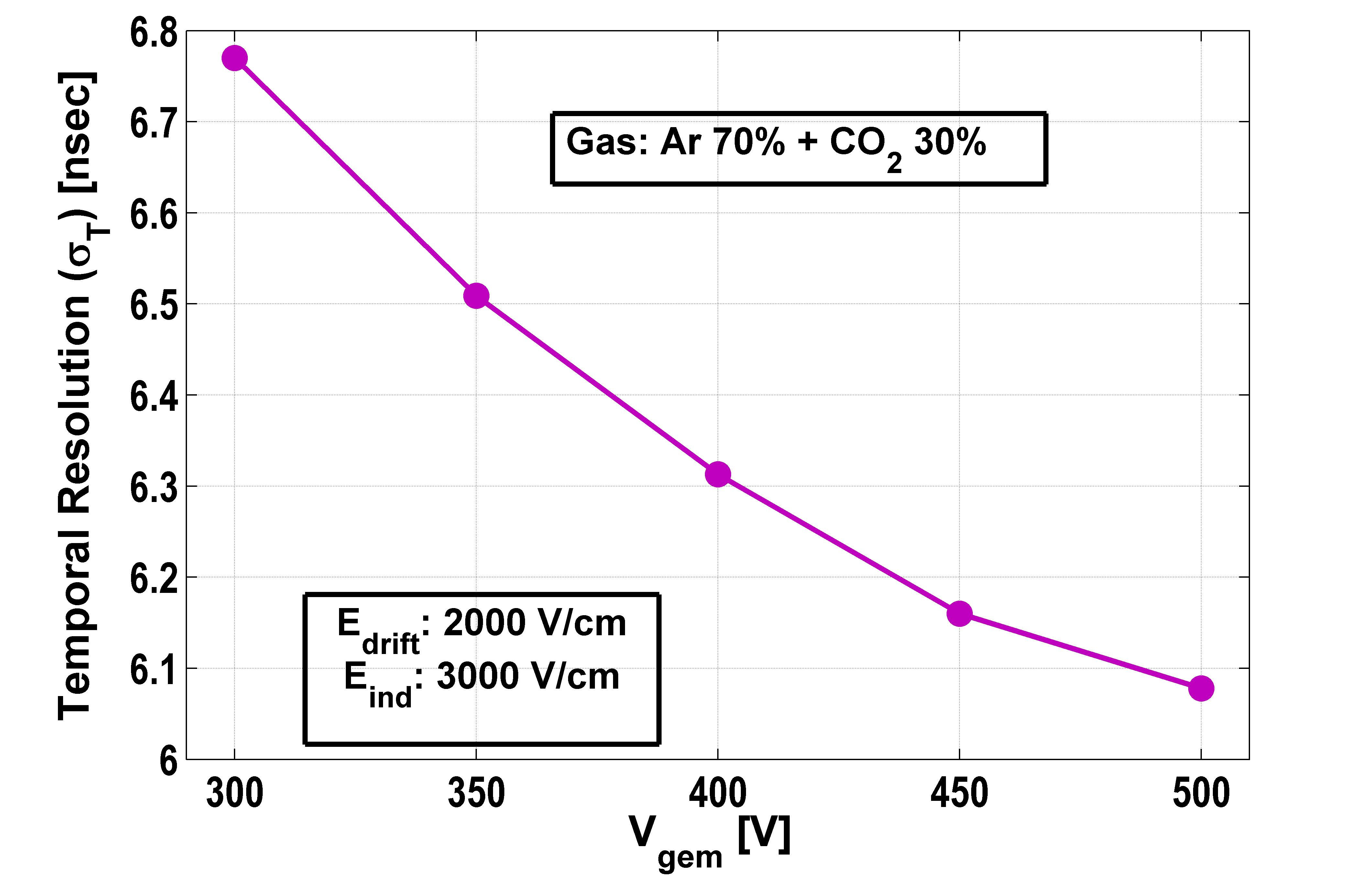}}
\subfigure[]
{\label{GEM-Hit-Ind}\includegraphics[scale=0.05]{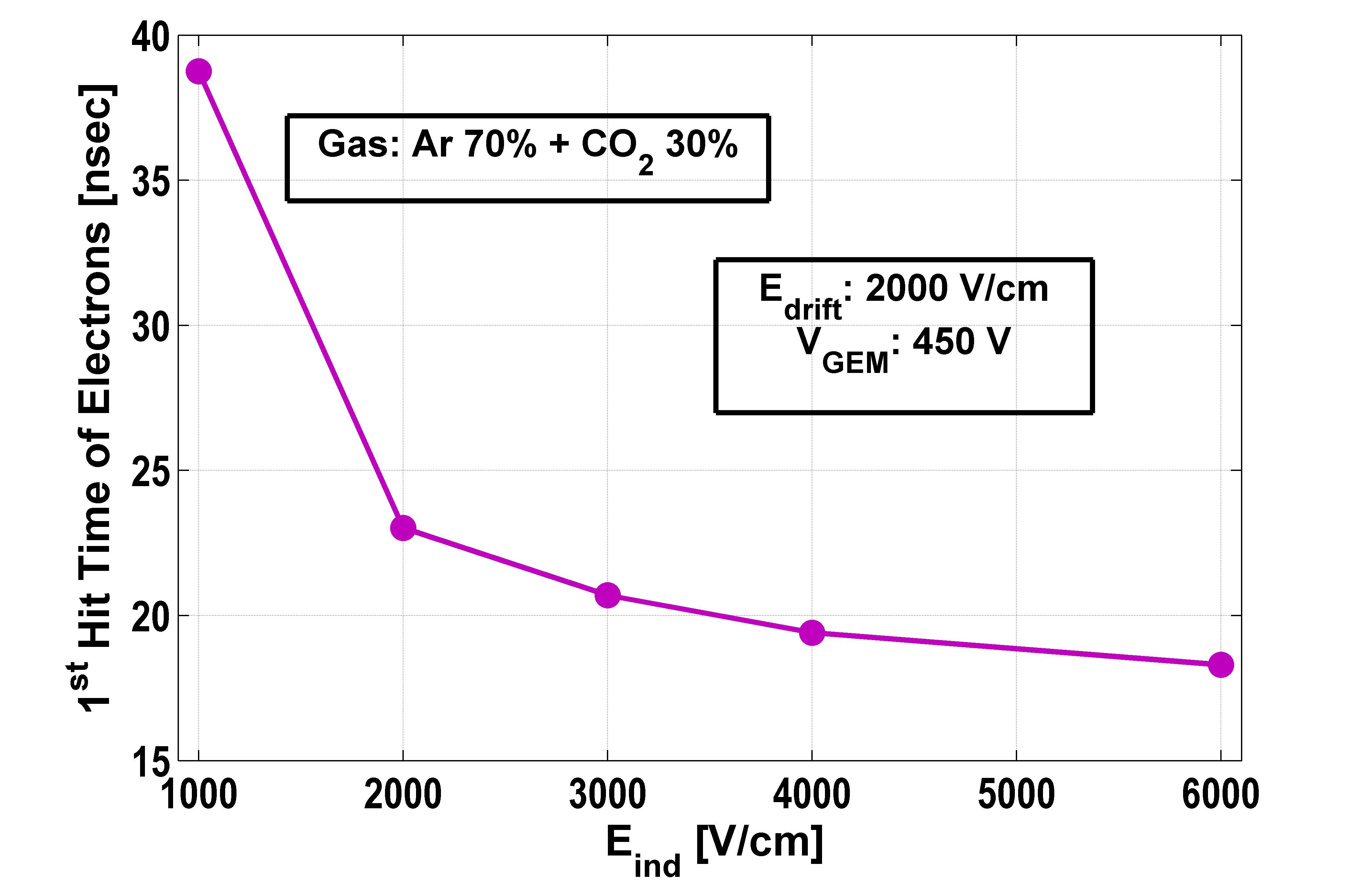}}
\subfigure[]
{\label{GEM-RMS-Ind}\includegraphics[scale=0.05]{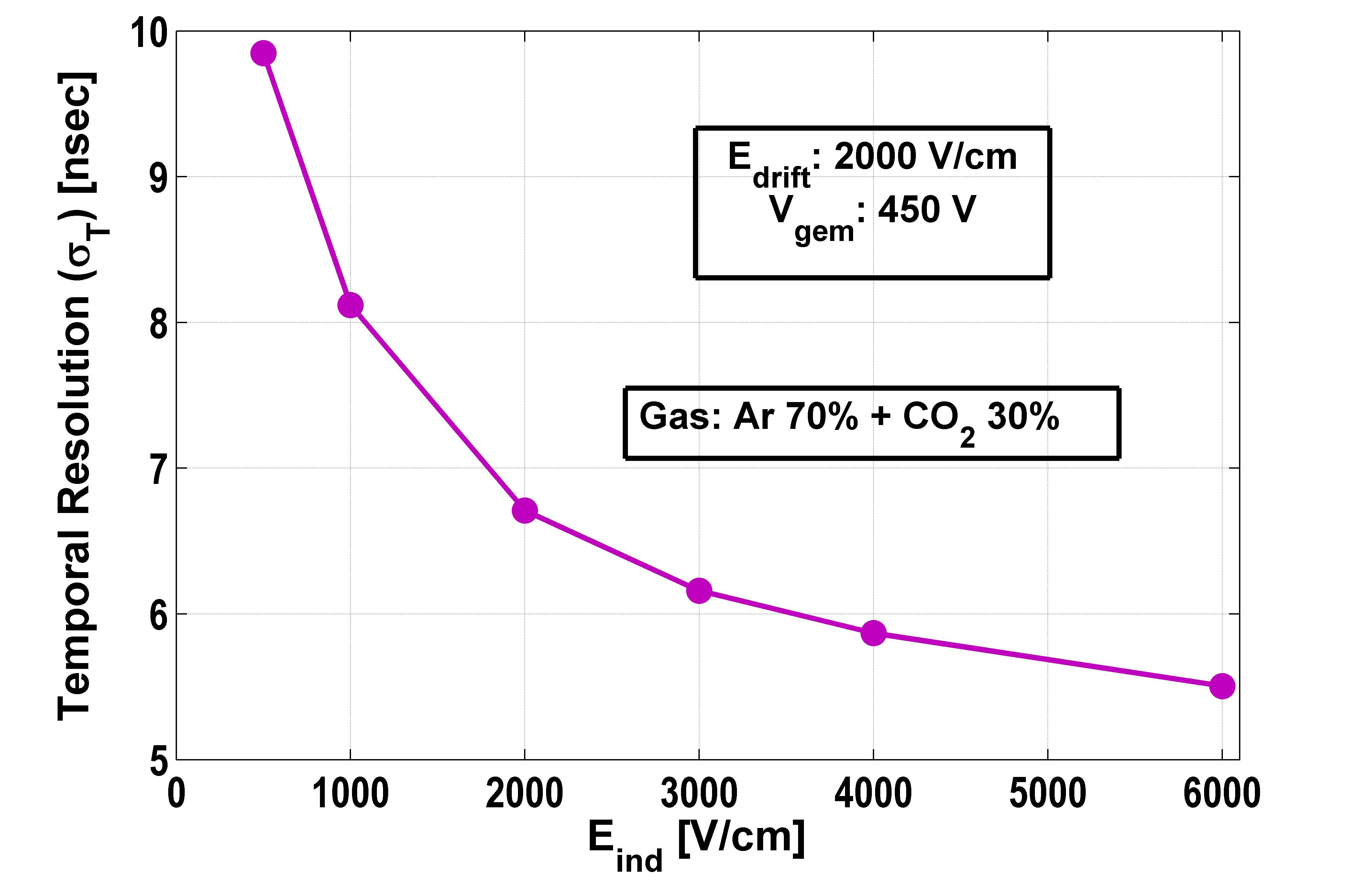}}
\caption{For a single GEM detector in $\mathrm{Ar}$-$\mathrm{CO_2}$ (${70:30}$) mixture, the variation of first hit time with (a) ${E_{drift}}$, (c) ${V_{GEM}}$, (e) ${E_{ind}}$. For the same detector, the variation of temporal resolution with (b) ${E_{drift}}$, (d) ${V_{GEM}}$, (f) ${E_{ind}}$. For (a) and (b), ${V_{GEM}} = 450~\mathrm{V}$, ${E_{ind}} = 3000~\mathrm{V/cm}$; for (c) and (d) ${E_{drift}} = 2000~\mathrm{V/cm}$, ${E_{ind}} = 3000~\mathrm{V/cm}$; for (e) and (f) ${E_{drift}} = 2000~\mathrm{V/cm}$, ${V_{GEM}} = 450~\mathrm{V}$.}
\label{GEM-Time3}
\end{figure}

To study the performance of GEM-based MPGDs, the temporal resolution of a single GEM detector and its dependence on the
electrostatic configuration have been numerically simulated at first.
The variation of temporal resolution with ${E_{drift}}$, ${V_{gem}}$ and ${E_{ind}}$ has been carried out in $\mathrm{Ar}$-$\mathrm{CO_2}$ ($70:30$) mixture.
Starting from the lower drift field, the resolution improves to a flat plateau.
At lower drift fields, larger diffusion is responsible for a larger drift time and worsening of resolution (Figure \ref{GEM-Hit-Drt} and Figure \ref{GEM-RMS-Drt}).
The variation with ${E_{ind}}$ and ${V_{gem}}$ reveal that the lower drift time and better temporal resolution can be obtained with  higher induction field (Figure \ref{GEM-Hit-Ind} and Figure \ref{GEM-RMS-Ind}) and higher GEM voltage (Figure \ref{GEM-Hit-Vgem} and Figure \ref{GEM-RMS-Vgem}), respectively.
This can be understood from the drift line plot.
At a higher induction field and high GEM voltage, the funneling of the electrons is such that the electrons have to travel smaller path to reach the readout plane.
Also, at these higher field values, the transverse diffusion is low, whereas the drift velocity has an increasing trend.
So the electrons take smaller time to reach the readout plane without much distortion.

\subsection{Triple GEM}

\begin{figure}[hbt]
\centering
\subfigure[]
{\label{TripleGEM-Time1}\includegraphics[scale=0.0475]{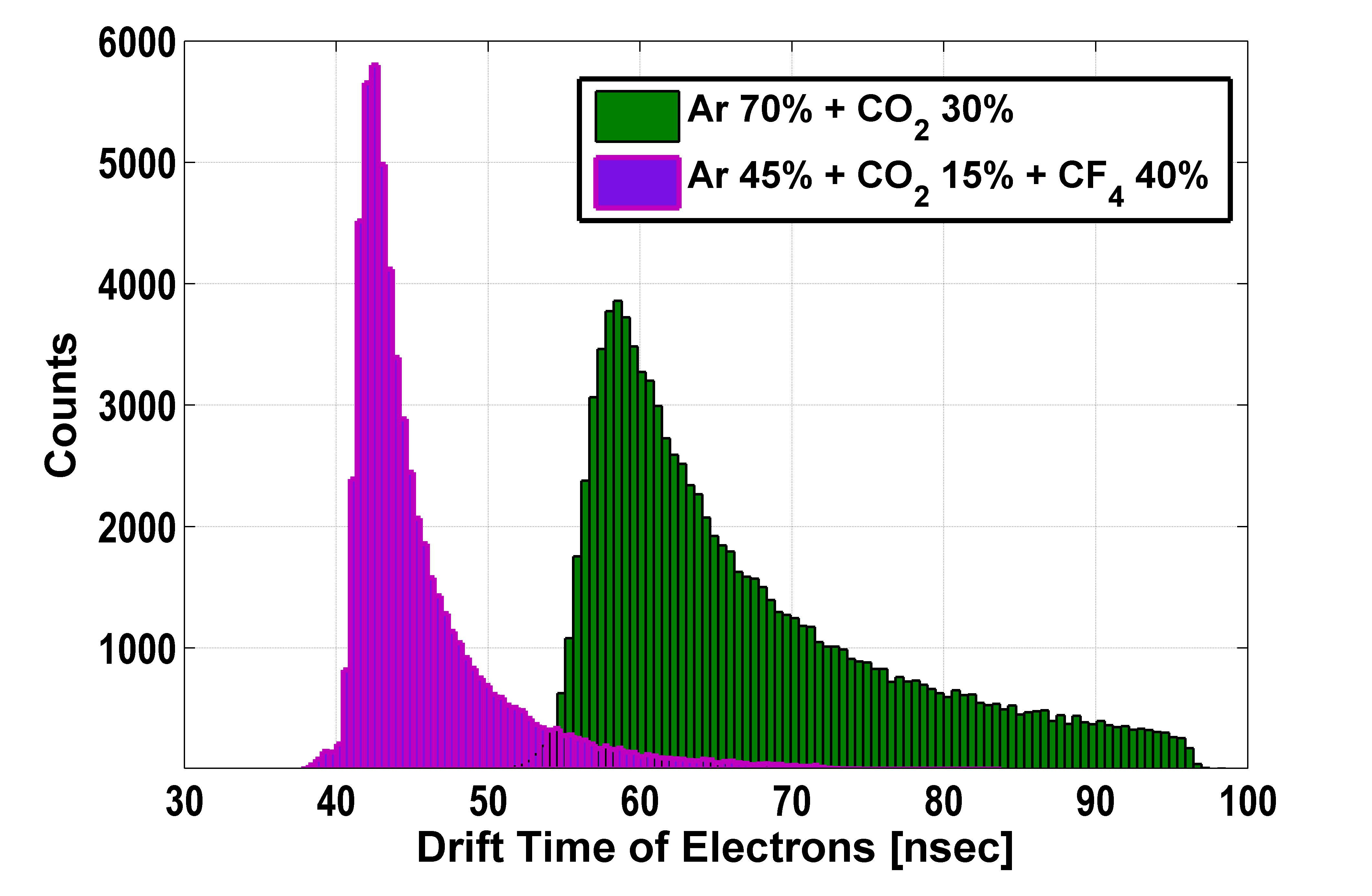}}
\subfigure[]
{\label{TripleGEM-Time2}\includegraphics[scale=0.05]{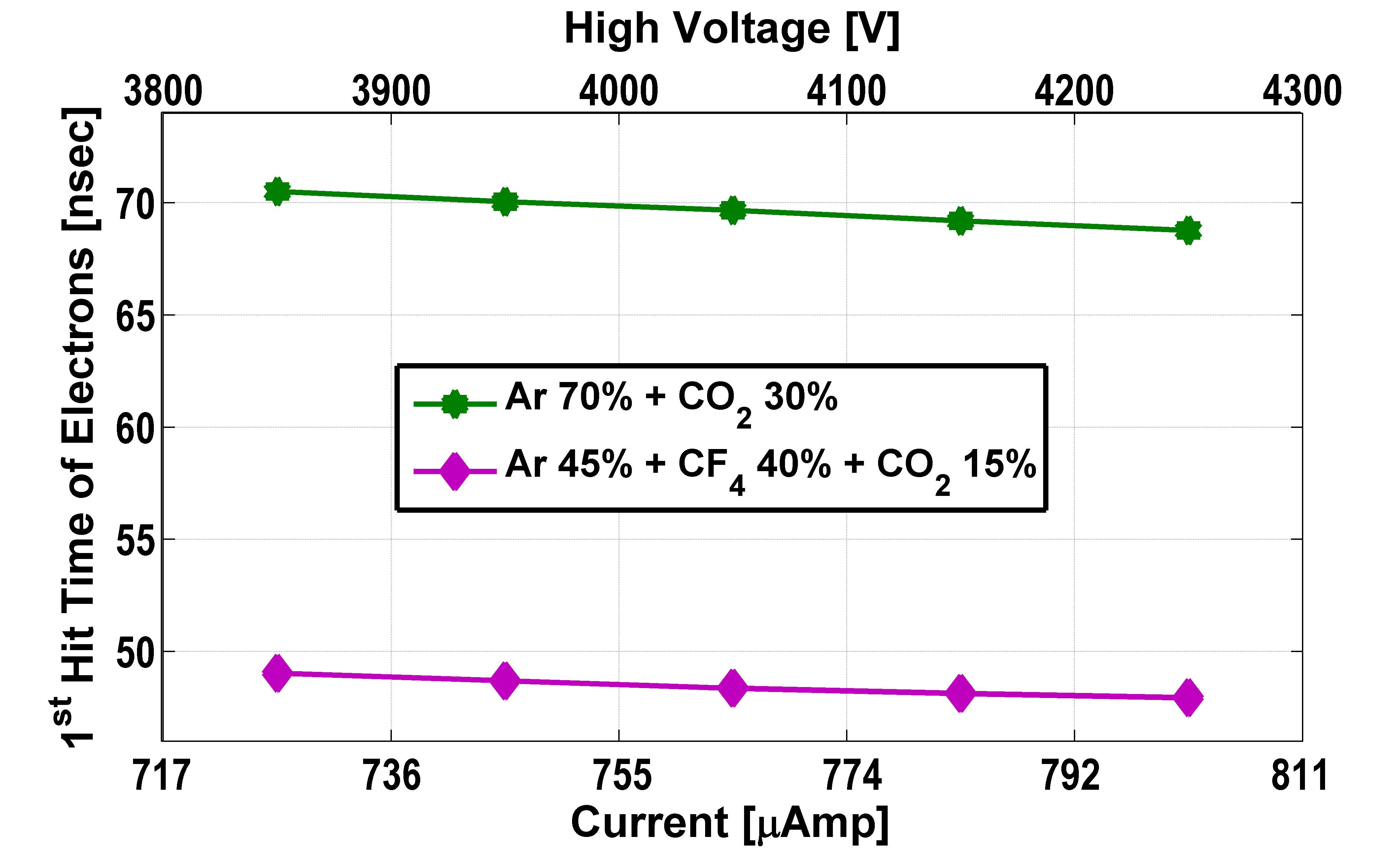}}
\subfigure[]
{\label{TripleGEM-Time3}\includegraphics[scale=0.05]{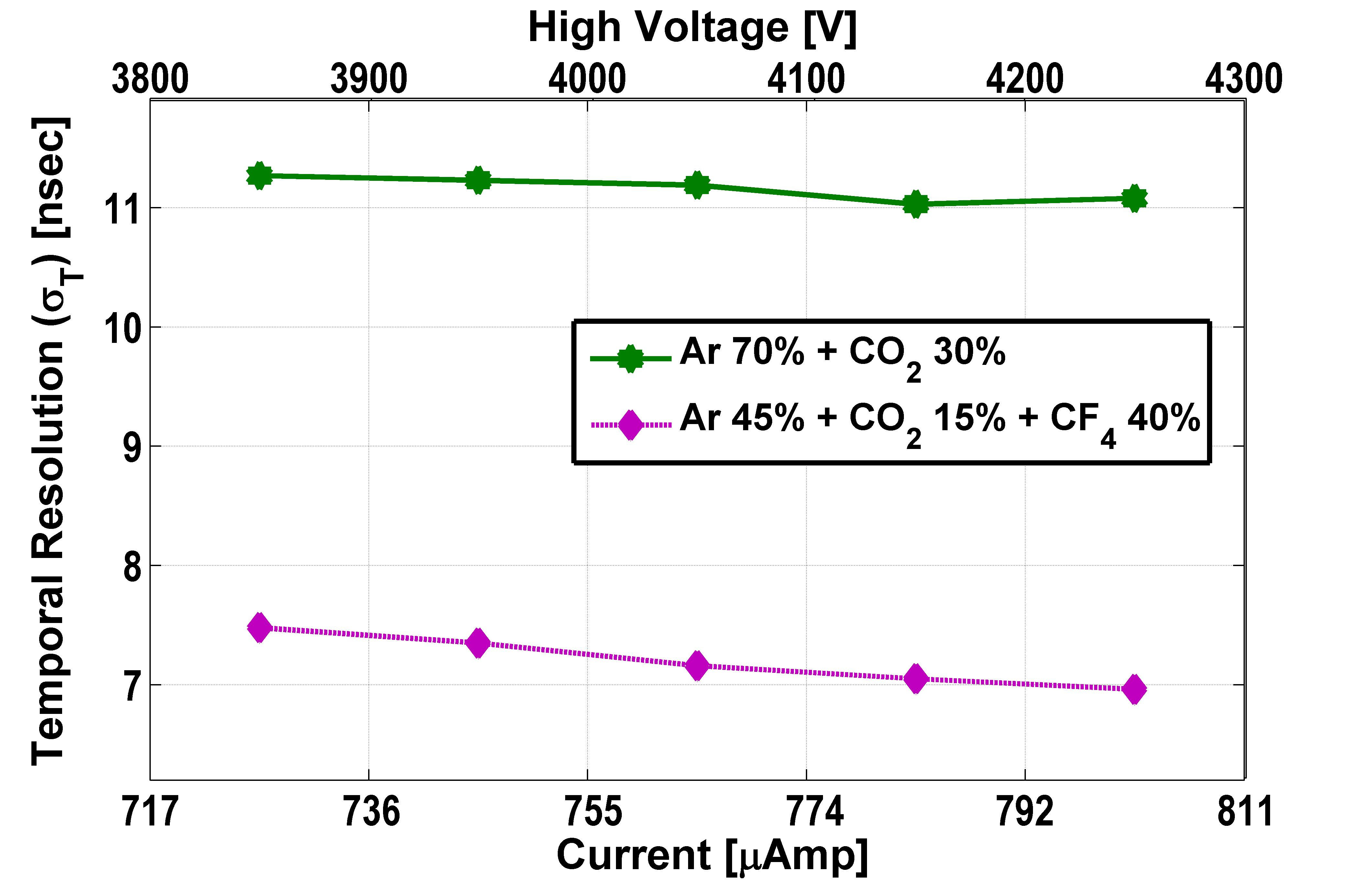}}
\caption{For a triple GEM detector: (a) time spectrum, the variation of (b) first hit time of electrons, (c) temporal resolution with current and applied high voltage in different $\mathrm{Argon}$-based gas mixture.}
\label{TripleGEM-Time}
\end{figure}

The time spectrum of the triple GEM detector as a second case of GEM-based detector, is shown in figure~\ref{TripleGEM-Time1} for two different $\mathrm{Argon}$-based gas mixture.
Due to higher drift velocity, the electrons in
$\mathrm{Ar}$-$\mathrm{CO_2}$-$\mathrm{CF_4}$ mixture take less time to hit the
 readout plane.
Also, the lower transverse diffusion coefficient in this mixture helps to
obtain a better temporal resolution.
It may be noted here that the numerical estimates ($\sim11~\mathrm{nsec}$ for
$\mathrm{Ar}$-$\mathrm{CO_2}$ and $\sim7~\mathrm{nsec}$ for
$\mathrm{Ar}$-$\mathrm{CO_2}$-$\mathrm{CF_4}$) are quite close to
experimentally measured values reported earlier \cite{TGEM1, Chapter6-TripleGEM2, Chapter6-TripleGEM4}.

The variation of the first hit time and the corresponding temporal resolution with applied high voltage for these two gas mixtures have been plotted in Figure \ref{TripleGEM-Time2} and Figure \ref{TripleGEM-Time3}, respectively.
As discussed in the case of single GEM, an increase of the drift field, hole voltage and the induction field reduce the time taken by the electron to reach the readout plane, as well as improve the temporal resolution.
For the triple GEM also, a larger value of the applied high voltage increases the field in the respective regions. So, at higher voltage, the drift time is less and the resolution is better.

\section{Modified Simulation Approach}

\begin{figure}[hbt]
\centering
\subfigure[]
{\label{RPC-Signal}\includegraphics[scale=0.6]{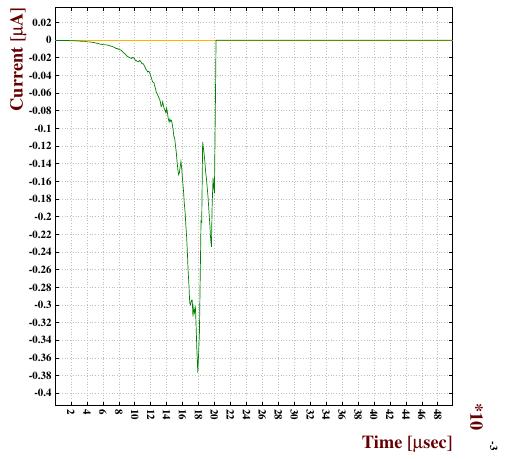}}
\subfigure[]
{\label{RPC-Time}\includegraphics[scale=0.8]{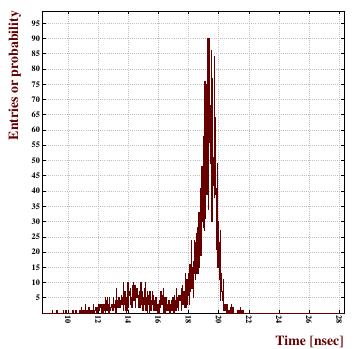}}
\caption{For RPC, (a) induced signal; (b) time spectrum}
\label{RPC}
\end{figure}

In the modified model, a threshold limit of detecting the
signal has been considered
which is related to the gain variance of the detector.
At first, the new approach has been applied on a RPC detector.
The detector dimension is listed in table~\ref{RPCdesign}.
High voltages of $\pm5800~\mathrm{V}$ have been applied in this detector to generate
the signal in $\mathrm{Freon}$-$\mathrm{Isobutane}$-$\mathrm{SF_6}$
(${95:4.5:0.5}$) gas mixture.
A typical induced signal for a cosmic muon track of $1~\mathrm{GeV}$
is shown in figure~\ref{RPC-Signal}.
For the present calculation, $0.1~\mu\mathrm{A}$ current at the
rising edge of the signal has been considered as a lower threshold.
For each track, the time to cross this threshold has been recorded
and the final spectrum looks like as shown in figure~\ref{RPC-Time}.
The r.m.s of this distribution has been found to $1.2~\mathrm{nsec}$
which is very close to the experimental value.
But this is only a preliminary calculation and further investigation is
going on before drawing any firm conclusion.
In future, this approach will be used to simulate the time resolution
for the MPGDs.

\section{Conclusion}

In this paper, an attempt has been made to establish an algorithm for
estimating the time resolution of an MPGD.
A comprehensive numerical study on the dependence of time resolution on
detector design parameters, field configuration and relative proportions of gas components has been made for a few MPGDs.
The simulated results have been compared with available experimental data and
the agreement between them is very encouraging.
Please note that, gas composition used for Micromegas and GEM are made different to compare them with the available
experimental data.
Thus, a comparison between these two MPGDs is not possible here.
The present work aims to accomplish a comprehensive characterization of the time
resolution of the MPGDs on the basis of numerical as well as experimental
measurements.
In addition to further improvement in the numerical work, development of a test
bench for studying the MPGDs invidually for their characteristics time
resolution and its dependence on the design parameters and gas composition has
been planned.

\section{acknowledgement}

This work has partly been performed in the framework of the RD51 Collaboration.
We wish to acknowledge the members of the RD51 Collaboration for their help and suggestions.
We thank our respective Institutions for providing us with the necessary facilities.

\end{document}